# RA V-Net: Deep learning network for automated liver segmentation

**Zhiqi Lee[1]，Sumin Qi[*], Chongchong Fan[2] and Ziwei Xie[3]**

[1]　Qu Fu Normal University，Cyberspace Security Institute，Qufu, Shandong，Chinese
[*]　Qu Fu Normal University，Cyberspace Security Institute，Qufu, Shandong，Chinese
[2]　Qu Fu Normal University，Cyberspace Security Institute，Qufu, Shandong，Chinese
[3]　Qu Fu Normal University，Cyberspace Security Institute，Qufu, Shandong，Chinese

**Email:**leezhiqi_edu@163.com,qixm@qfnu.edu.cn, 1425900035@qq.com,872035290@qq.com,

**Abstract**

Accurate segmentation of the liver is a prerequisite for the diagnosis of disease. Automated segmentation is an important application of computer-aided detection and diagnosis of liver disease. In recent years, automated processing of medical images has gained breakthroughs. However, the low contrast of abdominal scan CT images and the complexity of liver morphology make accurate automatic segmentation challenging. In this paper, we propose RA V-Net, which is an improved medical image automatic segmentation model based on U-Net. It has the following three main innovations. CofRes Module (Composite Original Feature Residual Module) is proposed. With more complex convolution layers and skip connections to make it obtain a higher level of image feature extraction capability and prevent gradient disappearance or explosion. AR Module (Attention Recovery Module) is proposed to reduce the computational effort of the model. In addition, the spatial features between the data pixels of the encoding and decoding modules are sensed by adjusting the channels and LSTM convolution. Finally, the image features are effectively retained. CA Module (Channel Attention Module) is introduced, which used to extract relevant channels with dependencies and strengthen them by matrix dot product, while weakening irrelevant channels without dependencies. The purpose of channel attention is achieved. The attention mechanism provided by LSTM convolution and CA Module are strong guarantees for the performance of the neural network. When testing, the evaluation index of the model has improved significantly, and the segmentation effect has a qualitative leap compared with the original model. In our experiments, we choose the publicly available datasets Lits2017, which includes 130 liver CT images of size 512*512 pixels. The accuracy of U-Net network: 0.9862, precision: 0.9118, DSC: 0.8547, JSC: 0.82. The evaluation metrics of RA V-Net, accuracy: 0.9968, precision: 0.9597, DSC: 0.9654, JSC: 0.9414. The most representative metric for the segmentation effect is DSC, which improves 0.1107 over U-Net, and JSC improves 0.1214.

## 1.Introduction

　　Automatic segmentation of organs in CT images is an application of computer-aided detection, which provides the basis for physicians to plan surgery. Realizing automatic segmentation of human organs has been a long-standing expectation of researchers. In recent years, with the research of deep learning, various segmentation metrics have gradually approached the application level. Organ segmentation is a prerequisite for finding the lesion region from the patient's CT image. The accuracy of organ segmentation will be directly related to the detection of lesion regions, so it is necessary to improve its accuracy.

The traditional image segmentation methods are: threshold segmentation, edge detection, region growing, Markov random field model, etc. Threshold segmentation (Bao et al., 2019; Tan et al., 2015; Zhang et al., 2010), which is characterized by simple implementation, small computational effort and stable performance. It is the most basic and widely used segmentation technique in image segmentation. By setting different thresholds, the target and background with gray value gap are segmented. Region growth (A and A, 2016; Ugarriza et al., 2009; Zhu et al., 2018), is the process of merging pixels or sub-regions into larger regions according to rules. The common algorithms include: region growing method with homogeneity, symmetric region growing method, and fuzzy connectedness method. Edge detection (Diaz-Pernil et al., 2013; Huang et al., 2009; Tabb and Ahuja, 1997; Wang and Oliensis, 2010), including serial edge detection, parallel edge detection. In recent years, methods based on surface fitting (Cai and Miklavcic, 2013), methods based on boundary curve fitting (Biswas et al., 2016), and methods based on Serial boundary finding have also been proposed. The basic idea is to use the existence of gray value disparity of pixels in different regions of edges to classify them. Markov random field (Singh et al., 2004; Xia et al., 2007), the objective function of the partitioning problem is determined according to the optimality criterion in statistical decision and estimation theory, and the maximum possible distribution satisfying these constraints is solved, thus transforming the partitioning problem into an optimization problem. Traditional segmentation methods are less effective, especially for medical images. Because medical images, like CT maps, have low contrast and similar grayscale values of organs, it is difficult to segment specific organs from a large number of organs and tissues. Traditional methods do not meet the requirements of accuracy in the field of medical image segmentation. Semi-automatic segmentation requires manual feature design and extraction. Therefore, the ability to express features is limited.

With the arrival of the era of big data, deep learning has made a big impact in various industries and fields. Especially in recent years, a large amount of training data has provided the development of automated segmentation of medical images, which has led to its rapid development. The network model of deep learning is getting higher and higher data of the evaluation index such as accuracy and recall, and the reliability are rising year by year. Alex -Net (Alex et al., 2012), proposed by Alex Krizhevsky et al. in 2012, it is influential for the image processing field until now. It used ReLU as the activation function to improve the nonlinear expression of neural networks, and also used dropout for the first time to ignore some neurons and shorten the training time while preventing overfitting of neural networks. 2015, Jonathan Long et al. proposed a fully convolutional neural network (Long et al., 2015). Unlike AlexNet for overall image classification, FCN can achieve pixel-level classification and thus semantic segmentation of images. FCN can also achieve end-to-end segmentation. By adding upsampling layers to the network structure, a segmentation result map of the same size can be obtained regardless of the size of the input image. This makes the applicability of FCN greatly improved, but there are still blurred and smooth results caused by upsampling, and the details of the image cannot be accurately perceived and segmented from the pixel level. In the same year, Olaf Ronneberger et al. proposed U-Net (Ronneberger et al., 2015), a deep network model for medical image processing, which has since become the baseline in medical image segmentation. The skip connection links the encoding and decoding layers on both sides of the U-shape, and passes the features of the encoding part directly to the decoding layer with the same image feature size without downsampling. This allows the image to retain more underlying features in the process of upsampling to recover size, and also allows features of different sizes to be fully fused. In 2015, He Kaiming et al. proposed Res Net (He et al., 2016) based on CNN network architecture, which cleverly solved two major problems of deep learning: gradient disappearance and overfitting by simple inter-layer jumping and feature passing. Subsequent researchers combined Res Net and U-Net, using residual block in the encoding module instead of the original encoding layer, and Res U-Net was created. Since then, deep neural networks are inseparable from lightweight residual networks. However, the segmentation results focus only on the relationship between neighboring pixels and ignore the spatial consistency of image features. In 2018, Ozan Oktay et al. proposed a neural network model Attention U-Net (Oktay et al., 2018) for pancreas segmentation, incorporating spatial attention mechanism based on U-Net network architecture. The network uses a feature weight matrix to supervise the upper layer features by the features of the lower layer, thus implementing the attention mechanism. The attention gate is added at the end of the jump connection, and the extracted features are weighted by the feature matrix, which allows the model to apply greater weights to the regions to be segmented. ROI (Region of Interest) receive higher weights, which in turn improves the accuracy of segmentation. In 2020, Guo et al. experimentally verified the effectiveness of the spatial attention mechanism (Guo *et al.*, 2020) and suggested that there is more room for improvement of the spatial attention mechanism. Mou et al. proposed CS2-Net (Mou et al., 2021) in 2021 to design an automated segmentation network with pixel-level spatial attention mechanism and channel attention mechanism co-existing. Two attention mechanisms are added to the jump connection part at the bottom of the model. The method of spatial attention is the feature matrix inner product. This allows the individual pixels of the image feature matrix to acquire spatial correlation. A similar approach is used for channel attention, where certain channels that are rich in feature information are acquired and the weights are elevated. To date, many new networks have been created with refreshing performance and structure, such as BCD U-Net (Azad et al., 2019), DeepLab (Chen et al., 2018), CE Net (Gu et al., 2019), U-Net++ (Zhou et al., 2018), DResUNet (Yu *et al.*, 2021a).

The main contributions of this paper are as follows: first, to address the complexity of segmentation targets in medical images, we propose CofRes Module, which can extract and preserve the maximum possible original image features by double-jump connection. Second, we introduce a channel attention mechanism, attempting to capture more image feature information in skip connection. We try to perform matrix transformation and activation of the image features. By enhancing the weights of some channels in the weight matrix, we achieve more accurate grasping of the image features. Finally, we propose the AR Module. Using the idea of residual networks, retaining the original image features and convolving the two sets of input image features using LSTM. Spatial attention mechanisms are added with the aim of enhancing the spatial relevance of the semantics to be segmented.

After detailed ablation experiments and performance tests, it was demonstrated that our proposed network incorporating three innovative ideas has achieved considerable results.

## 2. Architecture and method

### 2.1 V-Net

Structure of RA V-Net (Residual Attention V-Net) we proposed, based on the architecture of U-Net, and continues the encoding-decoding module commonly used in deep learning architecture in the field of medical image processing. The image pre-processing module is built. In this part, CT images are converted to HU (Hounsfield Unites) values and then windowed. It will be adjusted to the window value that is most suitable for segmenting certain organ. The essence of this operation is to highlight the outline of a specified area by adjusting the contrast and brightness for the purpose of enhancing the image.

In the encoding module, we designed CofRes Module with powerful feature extraction capability. During our experiments, we found that the feature extraction ability of Res U-Net improved significantly compared with U-Net, but it did not reach the expectation, so we decided to design our own feature extractor. In CofRes Module, advanced image features are obtained by adjusting feature channels and tighter intra-block connections. This design prevents gradient disappearance or gradient explosion while allowing more accurate extraction of image features by superimposing more convolution operations.

We introduce a channel attention mechanism (CA Module) at the skip connection. Since a large number of feature channels are stacked at the bottom of the model, we want to use the channel focus mechanism to obtain correlations between channels. CA Module will generate a feature dependency matrix by reshaping, transposing and inner-producting the feature matrix. In this matrix, channels with relevance are enhanced because they receive more weight. Channels without relevance are suppressed. Eventually, the image feature matrix and the feature dependency matrix are inner-producted. Therefore, some channels in the image feature matrix ware weighted.

In the decoder module, we tend to use upsampling combined with convolution to recover the dimensions of the image and incorporate a spatial attention mechanism, called AR Module. The concept of residuals is incorporated in our upsampling block to prevent overfitting. These channels are stacked heavily, so the convolution blocks that adjust the number of channels is set to reduce the computational effort of the model. Finally, the feature matrices passed by skip connections perform the reshaping together with the feature matrices output from the upsampling block. These feature matrices are fed into the LSTM for convolution after concatenation. The spatial attention mechanism is implemented by LSTM convolution to maintain the spatial correlation between pixels.

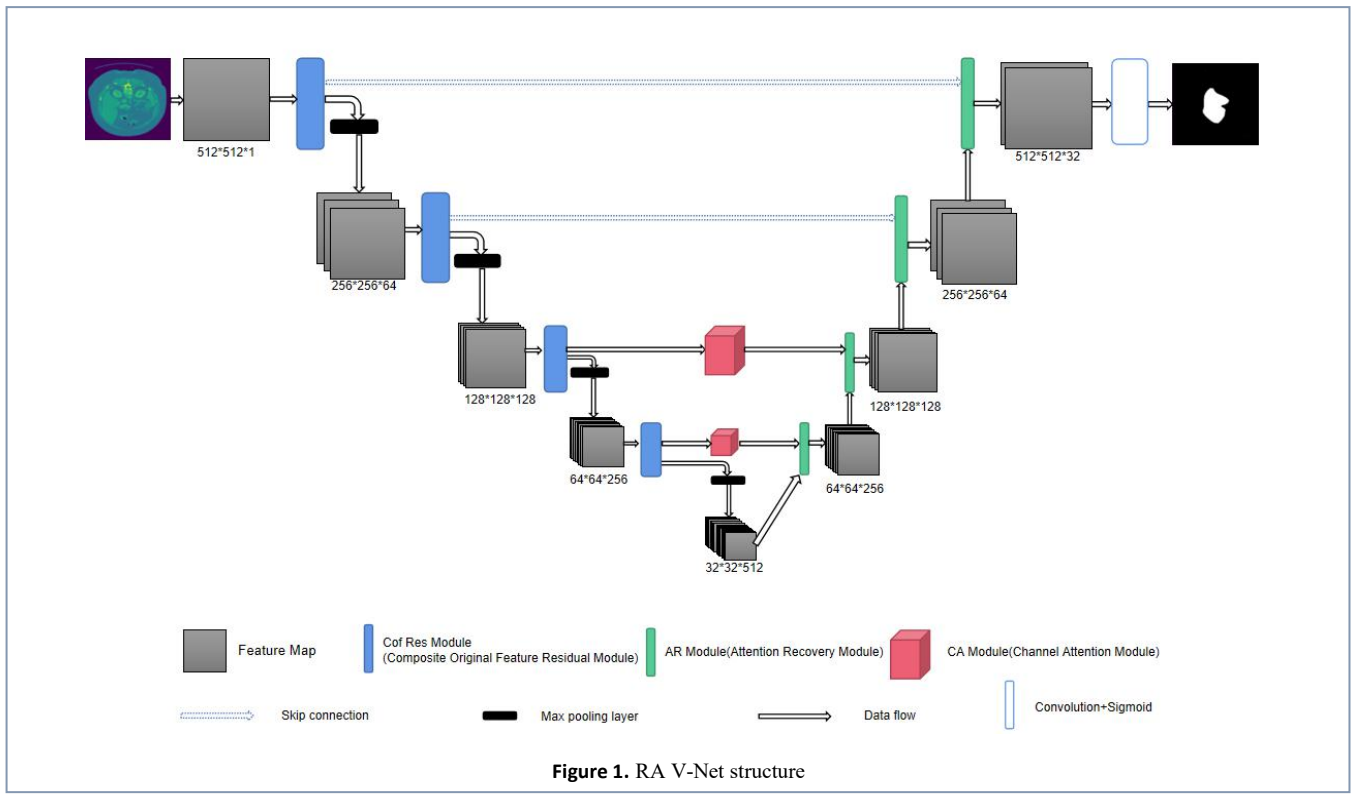

**Figure 1.** RA V-Net structure

The model we proposed has encoding part and decoding part , which is inspired by U-Net. However, compared with U-Net, our model replaces the bottom fully connected part with CofRes Module with stronger feature extraction capability. The network as a whole resembles a V-shape. Because of the combination of CofRes Module and two attention mechanisms we call it RA (Residual Attention) V-Net.

*2.2 Image pre-processing*

After loading the datasets, CT images needs to be converted to HU (Hounsfield Unites) values. The HU value is independent of the device used to take the CT image, and it is selected to represent different organs with different ranges of values. In addition, the HU value can represent the true density of the Ct map. In experiments, the range of HU values is usually large, which directly leads to low contrast of the image.

In order to improve the contrast and make the organs easier to be segmented, we use the Windowing method. Windowing is a processing method for CT maps that aims to highlight key structures in the image, and it is measured in HU values. Window Width (WW) and Window Level (WL) are two parameters in windowing. WW is a range. HU values within this range will be displayed. Units of HU values below the lower limit of WW are shown in black, and units of HU values above the upper limit of WW are shown in white. WL is a HU value, representing the middle value of all HU values in the window operation.

If WW increases, changing the gray shading of a certain HU value unit requires a greater change in density. As a result, more structures use approximate HU values, making the image contrast lower. As WL increases, the window is represented as white requiring a higher HU value. In summary, as WW increases, the image contrast decreases, and as WL increases, the overall image brightness decreases.

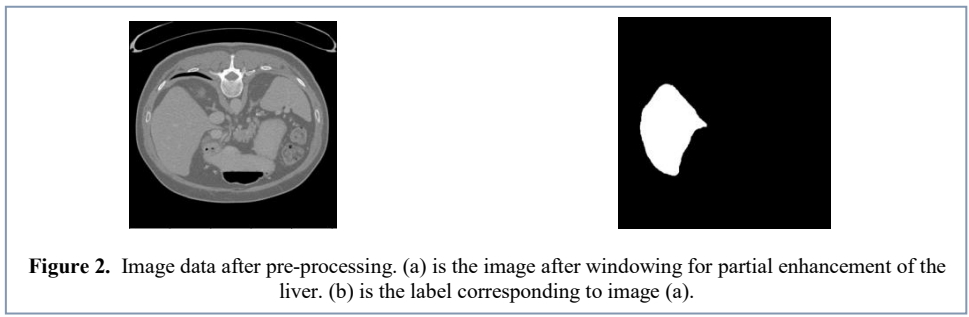

**Figure 2.** Image data after pre-processing. (a) is the image after windowing for partial enhancement of the liver. (b) is the label corresponding to image (a).

*2.3 CofRes Module*

CofRes Module (Composite Original Feature Residual Module), a residual module we propose, is applied to the encoding part of RA V-Net. Unlike the classical residual module, quadruple convolution and more dense intra-block connectivity are set in the CofRes module. The original feature matrix and the result of the convolution are summed to complete the retention of the input image features. In addition, by controlling the number of convolution kernels, shrinking the channels of the convolution operation and stacking multiple convolution results, the reduction of computation and number of parameters is achieved. Therefore, more convolution blocks and intra-block connections provide better performance for the CofRes module compared with the traditional residual blocks, such as overfitting prevention, feature extraction, accelerated operations, and feature retention. It is the core of our proposed model. A description of the CofRes Module structure is as follows:

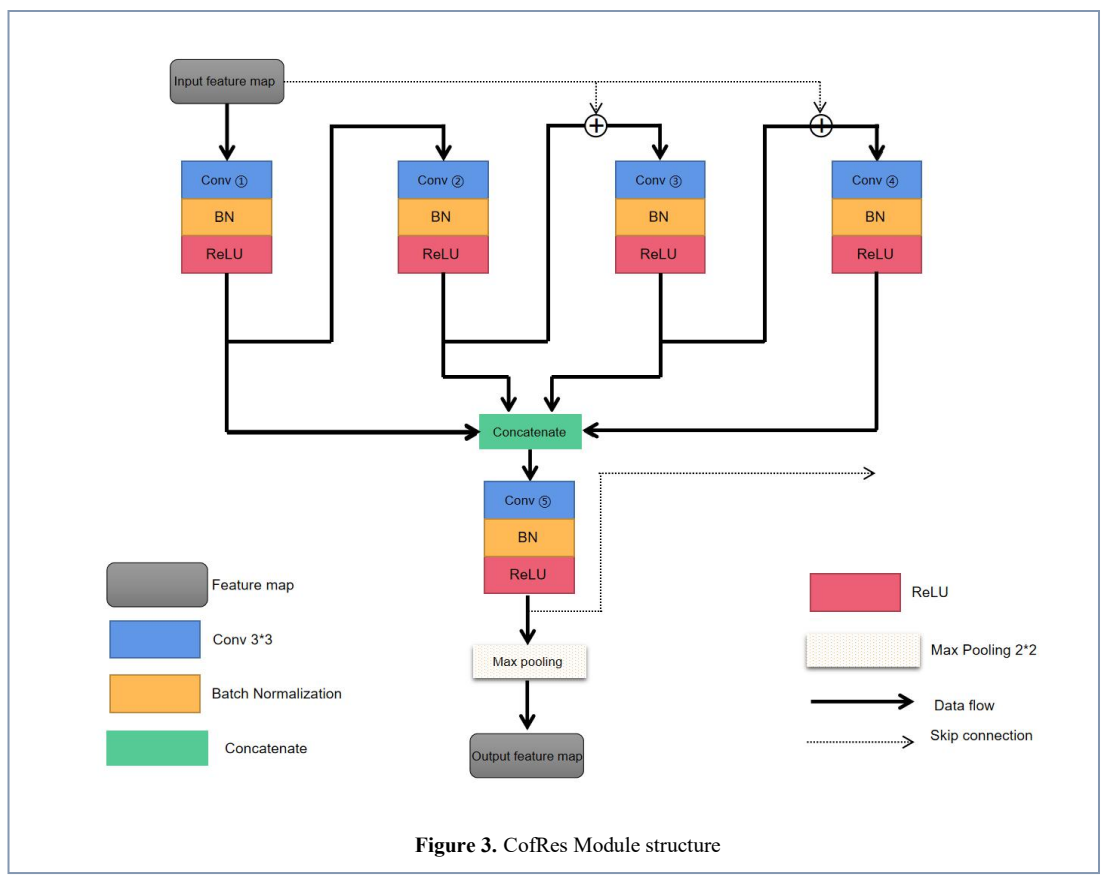

**Figure 3.** CofRes Module structure

In Figure 3, Conv①-④ are convolution operations with convolution kernel 3*3. The number of filters is set to 1/4 number of the target channels. The original feature matrix is summed with the convolved feature matrix to achieve preservation of the original image features. The convolved results are adjusted to the target number of channels by channel stacking. Conv⑤ is a 1*1 convolution operation that integrates the four convolution results. The feature matrix after fusion is used to skip connection and pass the feature matrix to the decoding module. Meanwhile, the input Maxpooling block shrinks the image size and is used as the input of the next CofRes Module.

Compared with the classical residual blocks, more convolution layers and denser intra-block connections are set in the CofRes module. Such a design can extract image features more accurately and prevent network degradation, gradient disappearance and gradient explosion. The features of the original image are retained in the convolution result by summation. While effective retention of original features provides guarantees for module performance, it allows the network to have deeper layers.

Using batch normalization and activation functions after each convolution can improve the nonlinear expressiveness of the convolution layers. In addition, the regularization of the inter-layer data stream and the nonlinear activation function allow us to stack more convolution layers into the network, improving the network's ability to extract features from the physical level. The number of channels is adjusted by convolution operations to integrate the extracted feature information.

*2.4 CA Module*

Channel attention module (CA Module) was introduced. In the deep features of an image, the channels mapped out by the matrix can be seen as category explicit corresponding objects, which are somehow associated with a specific semantic meaning. Therefore, we attempt to realize a feature matrix which enhances the representational power of image features for specific semantics through the dependencies mapped out between channels.

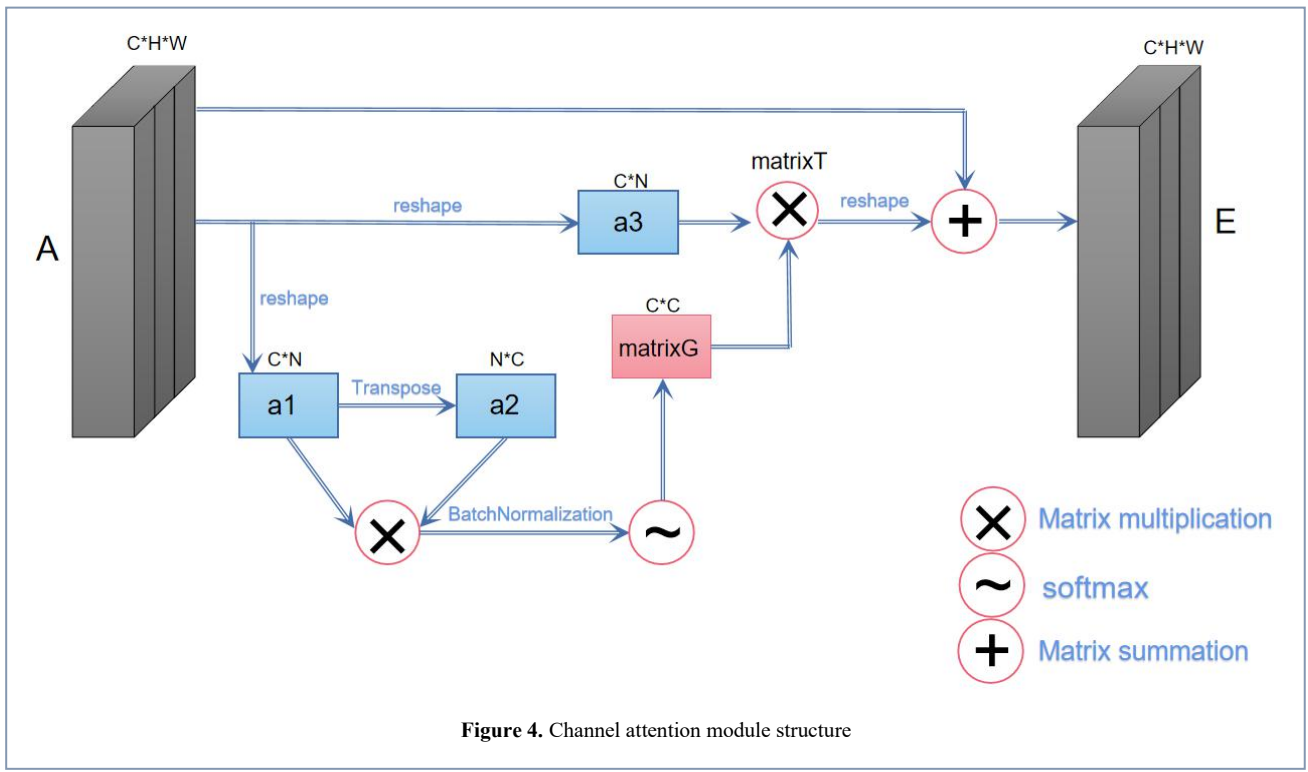

**Figure 4.** Channel attention module structure

In Figure 4, A is the depth image feature matrix input to CA Module from the encoding module, and E is the image features output from CA Module to the decoding module (C is the number of feature map channels, and H*W is the feature map size). a1 is the reshaping of the original feature A, and a2 is the transpose of a1. The channel dependency matrix G of size C*C is extracted by matrix multiplication, relying on the correlation within the feature map. After multiplying the original features and the channel dependency matrix T, the resulting matrix is called the channel weighting matrix T. Finally, the original feature matrix and the weighted feature matrix are used to obtain the output of CA Module. This is to prevent overfitting while strengthening the weights of the original features.

The channel dependence matrix G is extracted by matrix multiplication and softmax activation. In the feature matrix, two similar channels will promote each other. On the contrary, different channels will suppress each other. The more similar the semantic features of two image feature channels are, the higher the value of the channel feature matrix G at the corresponding position.

$$G_{(x,y)} = \frac{exp(W_x * W_y^T)}{\sum_{x'=1}^{C} exp(W_{x'} * W_y^T)} \tag{1}$$

where $G_{(x,y)}$ denotes the impact of the xth channel on the yth channel ($W_x$ represents a feature of 1*H*W. $W_y^T$ indicates a transpose of another feature of 1*H*W). Softmax is applied on the channel dependency matrix to enhance the discrimination between curvilinear structure and its background. After that, we reshape the original feature matrix A, then do matrix multiplication with $G_{(x,y)}$, multiply by the weight parameter β, and reshape again. Finally, the obtained feature matrix and the original feature matrix are added element by element to get the final feature map *E*.

The formulation of the attention mechanism in the channel injection is expressed as follows:

$$E_x = \beta \sum_{x=1}^{C} G_{(x,y)} \cdot A_{a3} + A_x \tag{2}$$

In Equation 2, β is initialized to 0 and learned gradually. And stating that each channel feature of the final output feature matrix is the result of a weighted sum of the channel weighting matrix T and the original feature matrix A. This is the reason why channel attention can work on the feature matrix.

*2.5 AR Module*

The AR Module (Attention Recovery Module) is presented. It is divided into two blocks: the upsampling block, and the LSTM block. The decoding part of our proposed model consists of AR Module.

In upsampling block, the dimensions of the feature matrix are recovered. In this process, residual ideas are incorporated. At the same time, the convolution operation to adjust the number of channels was set. Compared with the conventional upsampling block, the incorporation of residual ideas prevents overfitting. Meanwhile, adjusting the number of channels reduces the computational effort of the model.

In LSTM block, the data from upsampling block and skip connection input are reshaped. The adjusted data are sent to ConvLSTM. It provides spatial attention during convolution to capture the spatial correlation between pixels. The output data from ConvLSTM is convoluted, integrating the features, adjusting the number of channels and sent to the next AR Module.

The structural design of the AR Module is shown as follows:

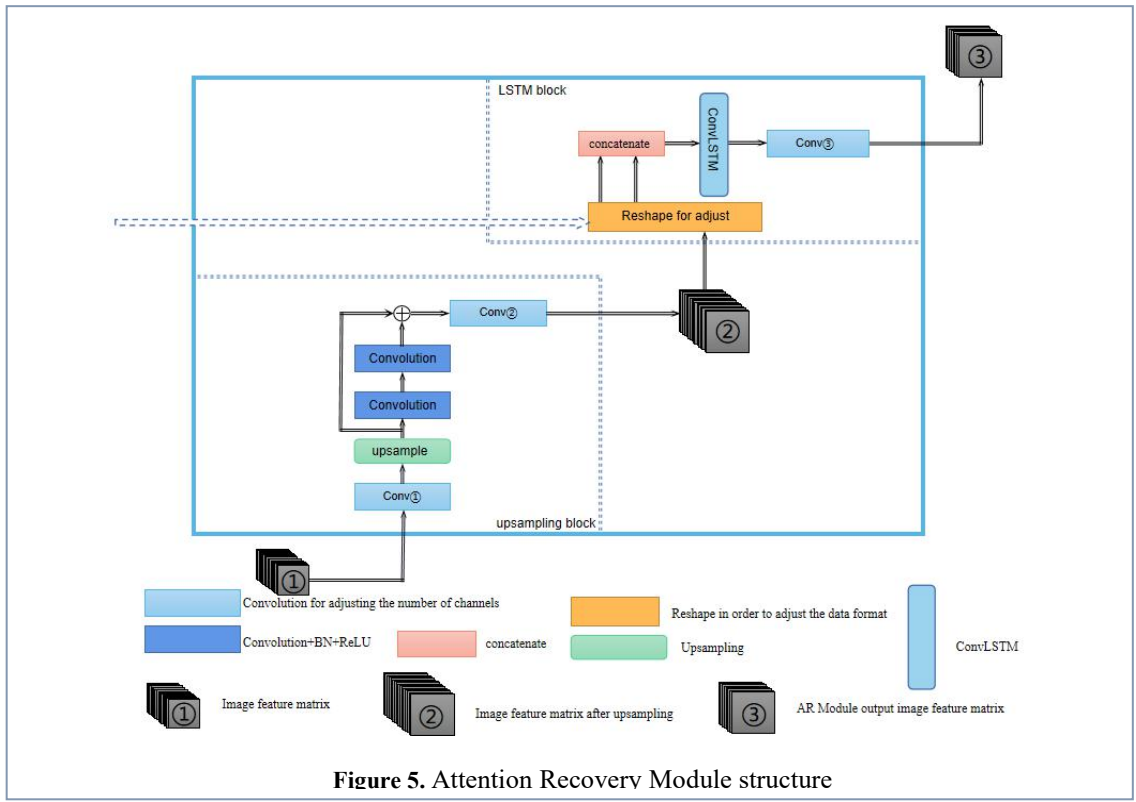

**Figure 5.** Attention Recovery Module structure

In Figure 5, the image feature matrix ① has the original size and the number of channels is the original number of channels. The feature matrix ② is the output of the input upsampling block of the feature matrix ①, its size is restored to the current size (twice the original size), and the number of channels is the original number of channels. Feature matrix ② is the input of LSTM block, feature matrix ③ is the corresponding output, the size of feature matrix ③ is the current size, the number of channels is the current number of channels (half of the original number of channels).

The number of channels is reduced to 1/4 of the original number of channels by Conv ①, after the feature matrix ① is input into the upsampling block. The purpose of this is to reduce the computational effort. After this operation is completed, the image size is restored. The classical residual block is then used to integrate the image features. Finally the number of channels is restored to the original number of channels by conv②. Output the feature matrix ②.

The data passed by the jump connection and feature matrix ② are input to the LSTM block for reshape. The format of the features is adjusted to (? , 1, H, W, C). Then, the two matrices are stacked in dimension 1. The format of the matrix becomes (? , 2, H, W, C). After that, the matrices are fed into ConvLSTM. A spatial attention mechanism is added to the two sets of data, relying on the spatial correlation between pixels to add weights to the corresponding positions. The format of the ConvLSTM output is (? , H, W, C). The final input conv ③ adjusts the number of channels for the feature matrix, shrinks to 1/2 of the original number of channels, and is then input to the next AR Module.

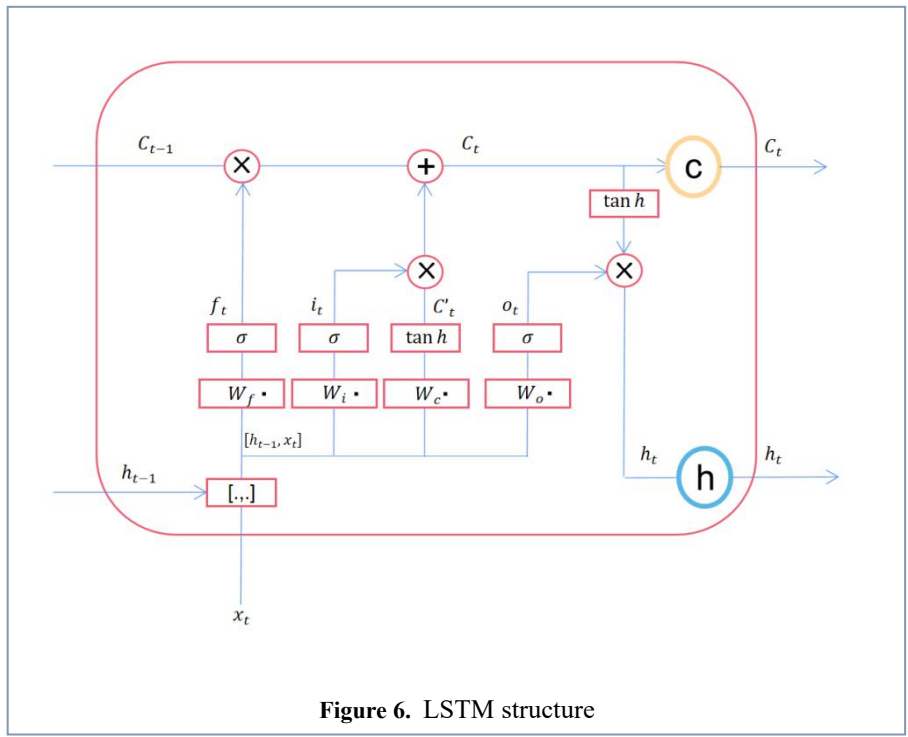

**Figure 6.** LSTM structure

In Figure 6, is the structure of the LSTM, which is powerful for processing temporal data. It shows surprising ability to capture spatial correlation of pixels when used to process image data. LSTM has three gates: input gate, output gate, and forget gate. It is worth noting that the weights within the LSTM computational unit are shared, with each layer of the LSTM sharing the same share of weights. In the LSTM cell, features are extracted by convolution, and candidate memory objects are generated using the three gates mentioned above and Tanh activation and Sigmoid activation. Among them, Tanh is chosen as the activation function of the candidate memory object because the output of Tanh is in -1~1, which is centered on 0. The gradient near 0 is large and the model converges quickly.

The cell state $c_{(t-1)}$ at the previous moment, first forgetting some unimportant information (determined by $f_t$) and adding some information (determined by $i_t$ and $c_t$) from the input at the current moment. The cell state is similar to a conveyor belt, running throughout the chain with only some few linear interactions. LSTM eventually generates a weight matrix that is enhanced with spatial attention for the input image features to perform site-specific attention enhancement, which in turn helps model segmentation.

*2.6 loss function*

The loss function is a tool, used to measure the difference between the predicted value of the model and the true value of the data, which is the direction of model training. There are many loss functions applicable to medical image segmentation, such as: Dice Loss, cross entropy, etc.

$$Loss_{dice} = 1 - \frac{2|x \cap y| + 1}{|x| + |y| + 1} \quad (3)$$

In equation 3, X is the prediction result of the model and Y is the Ground truth of the datasets.

Dice loss, calculate two sets and similarity measures, and 1 do the difference. The result is the dice loss, which calculates the proportion of the total number of pixels where the predicted image and the real image are 1 at the same time. Finally, the loss value is normalized to $[0 - 1]$, which is differentiable. Dice loss is a common loss function in medical images.

$$Loss_{bce} = -\frac{1}{N}\sum_{i=1}^{N}\{y_i * \log[p(y_i)] + (1 - y_i) * \log[1 - p(y_i)]\} \quad (4)$$

In equation 4, $y_i$ is the label (0 or 1) and $P(yi)$ is the probability that the model predicts a pixel point to be 1.

The BCE loss, which is the prediction result, is first normalized to $[0 - 1]$ by the sigmoid function, and then the bce loss formula is used to calculate the loss value. It is commonly used for binary classification, while image segmentation is a semantic binary classification task.

**Table 1.** Difference performance due to adaptation of different loss functions

| Loss function | Acc | Pre | DSC |
|---|---|---|---|
| Dice Loss | 0.9931 | **0.9545** | **0.9368** |
| BCE Loss | 0.9933 | 0.9336 | 0.9144 |

Dice loss and BCE loss, which are commonly used loss functions in image segmentation (semantic classification) tasks. We evaluate metrics differently by experimenting with different loss functions in the same model. After ten experiments, we averaged the evaluation metrics to obtain TABLE I. It can be seen that in the dataset LiTS2017, it is more suitable to use dice loss as the loss function to train the network parameters.

*2.7 Performance evaluation metrics*

*2.7.1 ACC*

Accuracy, the metric we referenced most often, is used to indicate the percentage of samples with correct predictions over all sample data.

Accuracy is a very common performance metric in deep learning, but it does not fully represent the performance of a model. For example, when the background is black and the target to be segmented is small, the end-to-end model predicts the segmentation result as all-black while ensuring the same size of the input and output images, and the accuracy of the model alone is high at this time, but such a model does not have the segmentation performance in the real sense.

$$\text{Accuracy} = \frac{TP+TN}{TP+TN+FP+FN} \quad (5)$$

*2.7.2 Pre*

Precision, indicating the proportion of all samples predicted to be positive by the model that are true positives.

$$\text{Precision} = \frac{TP}{TP+FP} \quad (6)$$

*2.7.3 Dice Similarity Coefficient (DSC)*

The most convincing evaluation metric for the semantic segmentation process, the pixel-level similarity between the segmentation results and Ground Truth is evaluated.

$$\text{DSC}(\text{result}, \text{target}) = \frac{2|\text{result} \cap \text{target}|}{|\text{result}| + |\text{target}|} \quad (7)$$

*2.7.4 Jaccard Similarity Coefficient (JSC)*

Indicates the relationship between the segmentation result and Ground Truth: the intersection of the two as a proportion of the concurrent set of the two. It has the same meaning as the evaluation index IOU (intersection over union), which often appears in papers in recent years. JSC and DSC, both of which are commonly used to evaluate the model performance and pixel-level ground truth of segmentation results.

$$\text{JSC}(\text{result}, \text{target}) = \frac{|\text{result} \cap \text{target}|}{|\text{result} \cup \text{target}|} \tag{8}$$

## 3. Experiments and conclusions

*3.1 Experimental environment*

For the training phase of the experiment, we used a Linux server with Ubuntu installed, two NVIDIA TESLA V100 32G compute cards, Anaconda3, Python 3.6, and TensorFlow-GPU 2.3.0. Windows and NVIDIA RTX2070 graphics cards were used for the testing phase.

In the model building part, dice is selected as the training loss, the learning rate is set to 1e-4, batch size is set to 1. In the callback function, we take the loss value 5e-4 as the function exit and Epoch is set to 500.

*3.2 Datasets*

The datasets we chose is the publicly available LiTS (2017) liver tumour segmentation challenge datasets. In this datasets, there are 130 enhanced CT abdominal scans in train set and 70 CT abdominal scans in the test set. We intercepted some of these data for experiments and tests. We used 80 percent of the data for training and 20 percent for testing. When training the model with keras, we applied the integrated function of separating the datasets, using 80 percent of the training set as training and 20 percent as validation.

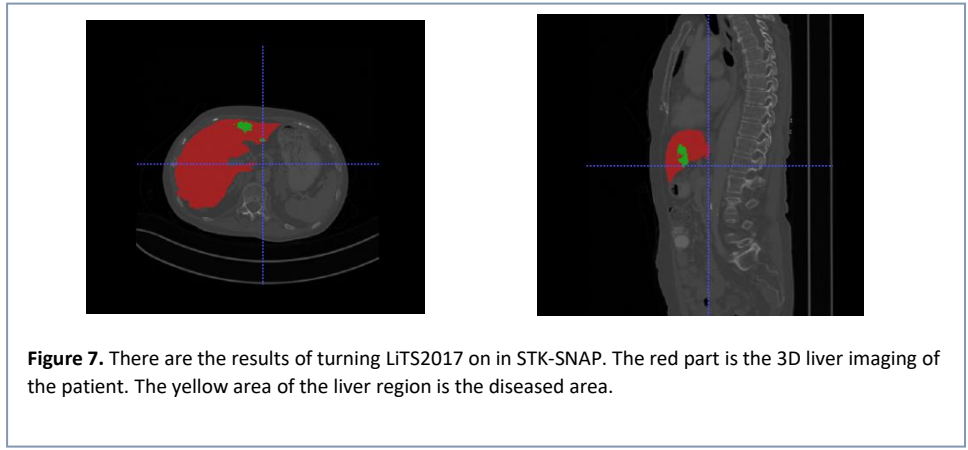

**Figure 7.** There are the results of turning LiTS2017 on in STK-SNAP. The red part is the 3D liver imaging of the patient. The yellow area of the liver region is the diseased area.

*3.3 Performance test*

*3.3.1 Encoding module performance*

The models involved in testing are: U-Net (Ronneberger *et al.*, 2015), Res U-Net, and CofRes U-Net (only U-Net equipped with CofRes Module as the encoding module). U-Net is the backbone model. Res U-Net is an initial innovation of U-Net with a encoding module combined with residual blocks. The purpose of this is to make the network have better feature extraction ability and also can have better resistance to overfitting. CofRes U-Net is to replace the encoding module of Res U-Net with CofRes Module, which is used to test the performance of CofRes Module. We set the loss 1e-3 as the training exit in the callback function, and the performance was evaluated uniformly after the models reached this level.

After complete training, the performance evaluation results on the test set are as follows:

**TABLE 2.** During encoding module testing, the performance metrics of different models, Accuracy, Precision, Dice, Jaccard similarity score

| Method | Acc | Pre | DSC | JSC |
| --- | --- | --- | --- | --- |
| U-Net | 0.9862 | 0.9118 | 0.8547 | 0.82 |
| Res U-Net | 0.988 | 0.9221 | 0.88 | 0.83 |
| CofRes U-Net | **0.9919** | **0.9377** | **0.9402** | **0.89** |

The results in **TABLE 2.** show that the performance of Res U-Net is significantly better than that of U-Net. this indicates that the residual block is unquestionably suitable for the encoding module. the performance of CofRes U-Net steadily exceeds that of Res U-Net, which shows that CofRes Module performs better than Res block.

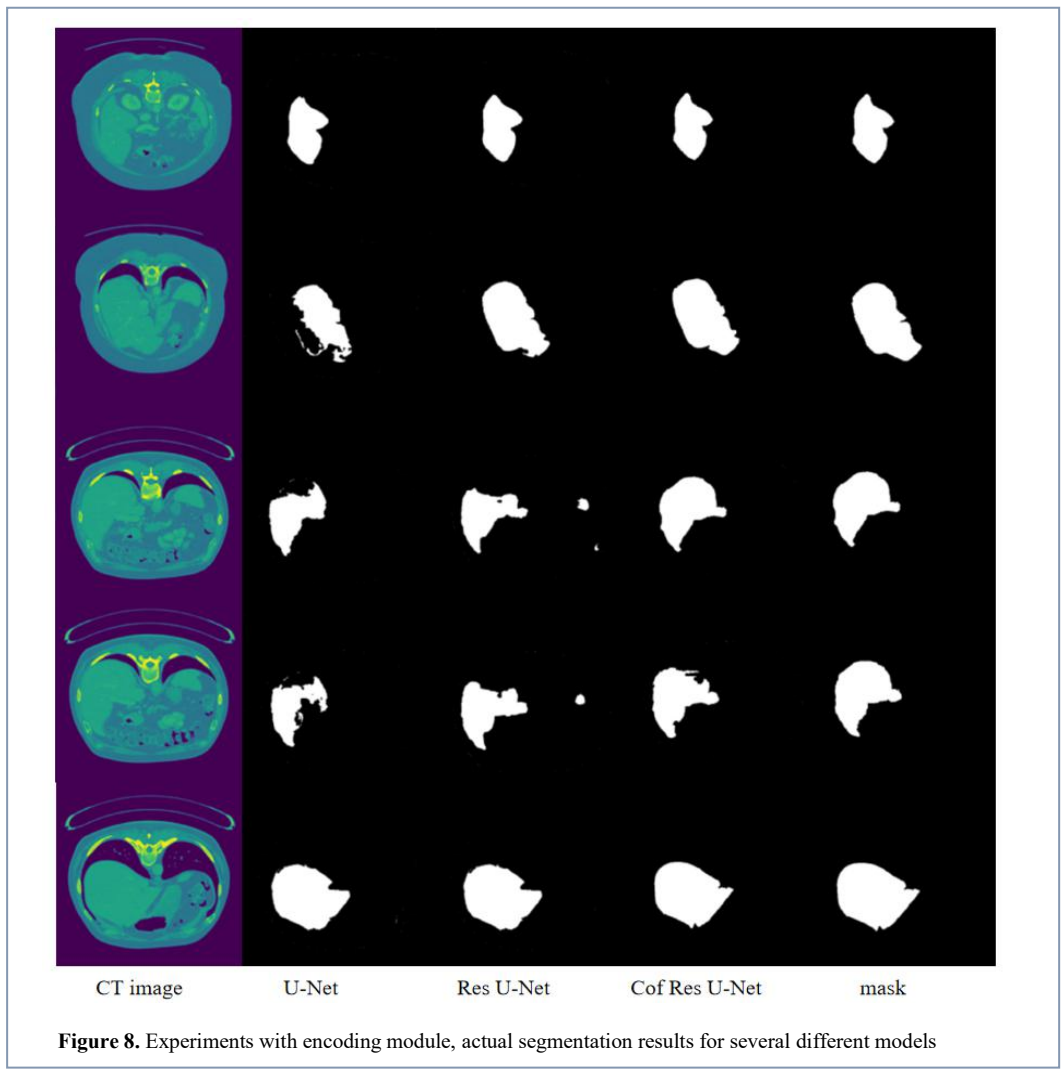

**Figure 8.** Experiments with encoding module, actual segmentation results for several different models

From the segmentation effect in Figure 8, we can see that although the CofRes U-net has more obvious performance advantages compared with other networks, the segmentation effect is still not good enough to meet our expectation of the model.

*3.3.2 Attention mechanism and decoding performance*

We expect to test the performance of the model in the attention mechanism and decoding module through experiments, mainly testing the model's ability to retain the original image features. Among the networks participating in the test, U-Net uses a simple jump connection and upsampling block in the decoding part, and SA U-Net adds a spatial attention mechanism in the jump connection based on U-Net to help the model refine the spatial attention matrix to enhance the spatial relevance of semantic segmentation. SA U-Net can be used as a control to test the performance of spatial attention and channel attention in our proposed model. In the architecture of BCD U-Net, the encoding part is the same as U-Net, and in the decoding part newly adds LSTM to implement the attention mechanism. To highlight the effect of CA Module and AR Module on the model performance, we participate in the test with BCD U-Net and test it against our model. To demonstrate the impact of LSTM on model performance, we performed a comparative test of BCD U-Net and U-Net (in this experiment, we removed the CofRes Module from the model and replaced it with the encoding module of U-Net for the purpose of controlling the variables. This makes the four models participating in the experiment identical in the encoding module.).

**Table 3.** During decoding module testing, the performance metrics of different models, accuracy, precision, Dice, Jaccard similarity score

| Method | Acc | Pre | DSC | JSC |
| --- | --- | --- | --- | --- |
| U-Net (Ronneberger *et al.*, 2015) | 0.9862 | 0.9518 | 0.8547 | 0.82 |
| SA U-Net (Guo *et al.*, 2020) | ------- | 0.91 | 0.9105 | **0.86** |
| BCD U-Net (Azad *et al.*, 2019) | 0.9879 | 0.87 | 0.8959 | 0.85 |
| Ours (without CofRes module) | **0.9903** | **0.9387** | **0.9214** | **0.86** |

The data in **Table 3** shows that the performance of SA U-net with the spatial attention module has a significant advantage over U-Net without this module. From this, it can be inferred that there is a direct positive correlation between preserving the spatial relevance of segmentation semantics and the performance of semantic segmentation. The performance of BCD U-Net using LSTM as spatial attention is also inferior to SA U-Net, indicating that the former attention performance is inferior to the attention mechanism generated by matrix computation. However, the attention mechanism in SA U-Net requires complex operations to generate. This will have a large impact on the computational speed of the model during training, and in addition will increase the computational pressure of the model. This is the reason why we did not choose it. Ultimately, the performance of SA U-Net is still lower than our proposed one (even though it does not carry CofRes Module), which suggests a combination of channel attention mechanisms and spatial attention mechanisms, a combination that provides strong performance guarantees.

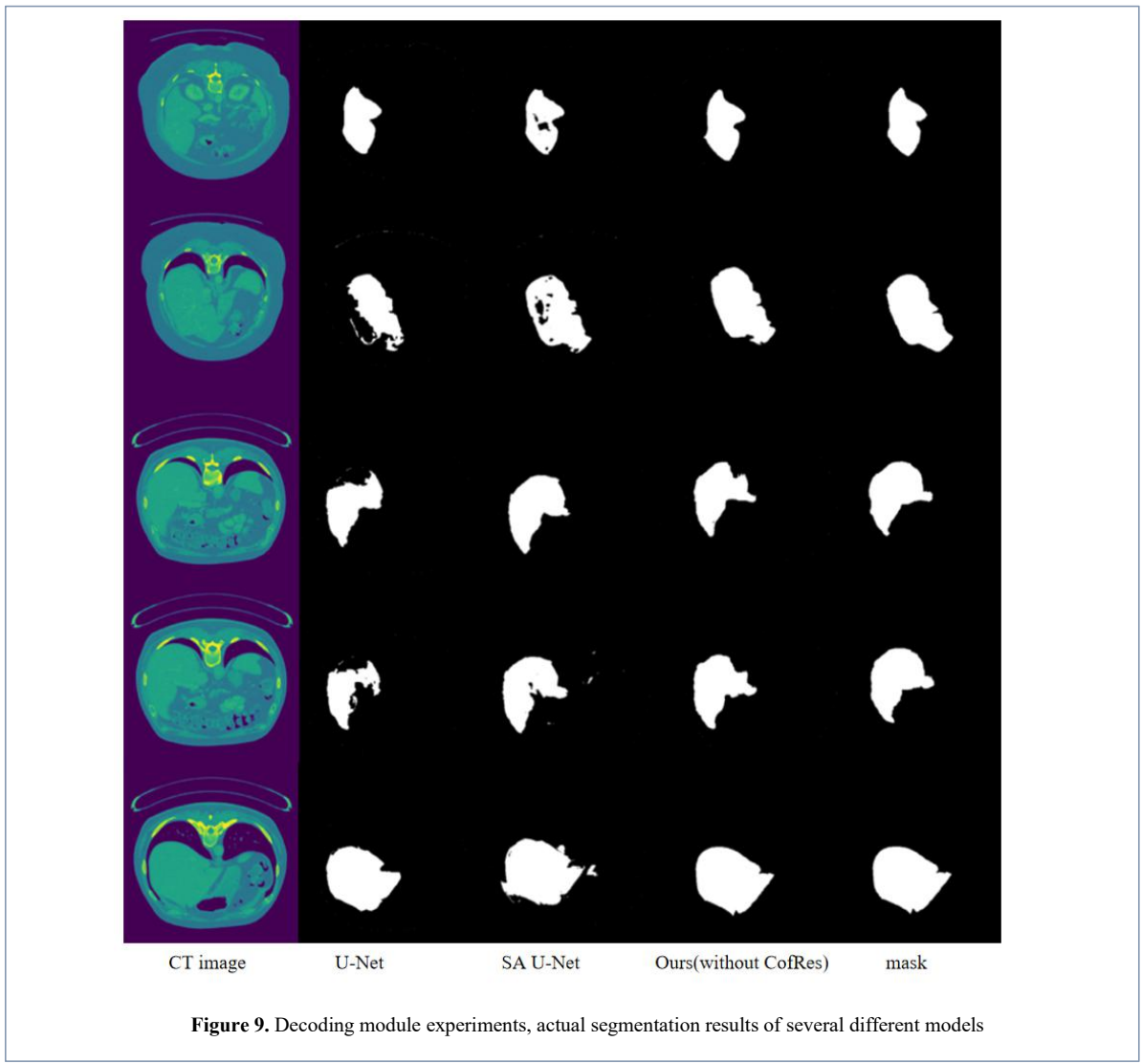

**Figure 9.** Decoding module experiments, actual segmentation results of several different models

For the decoding module, our ablation experiments also proceeded very well. The encoding part is the same as U-Net, and the decoding part uses the model performance of CA Module and AR Module. Although it does not reach our expectation, the segmentation results of our model are closer to the labels as seen in Fig 9. This illustrates the need to innovate on both encoding and decoding modules.

*3.3.3 Overall module performance*

After testing the functionality of our model in modules, we also test the model as a whole and compare the performance with classical neural network models and neural network models born in recent years, respectively. Here is a comparison of the performance results.

Table 4. Performance metrics of different models segmenting liver on LiTS2017, Accuracy, Precision, Dice, Jaccard similarity scores.

| Method | Acc | Pre | DSC | JSC |
| --- | --- | --- | --- | --- |
| U-Net (Ronneberger *et al.*, 2015) | 0.9862 | 0.9118 | 0.8547 | 0.82 |
| SA U-Net (Guo *et al.*, 2020) | -------- | 0.91 | 0.9105 | 0.86 |
| BCD U-Net (Azad *et al.*, 2019) | 0.9879 | 0.87 | 0.88 | 0.85 |
| Res U-net | 0.988 | 0.9221 | 0.88 | 0.83 |
| Mo (Mo *et al.*) | -------- | 0.9315 | 0.9369 | 0.8749 |
| CSDResU-Net (Yu *et al.*, 2021b) | -------- | 0.9224 | 0.9400 | -------- |
| ULSM-Net (Zhang *et al.*, 2021) | -------- | **0.9635** | 0.9518 | -------- |
| CS2 Net (Mou *et al.*, 2021) | 0.9923 | 0.9512 | 0.9587 | 0.9335 |
| Ours (without CofRes Module) | 0.9903 | 0.9387 | 0.9214 | 0.86 |
| Ours (CofRes U-Net) | 0.9919 | 0.9377 | 0.9402 | 0.89 |
| Ours (RA V-Net) | **0.9968** | 0.9597 | **0.9654** | **0.9414** |

We also collected and plotted the convergence of different models in terms of loss function and precision during the training process

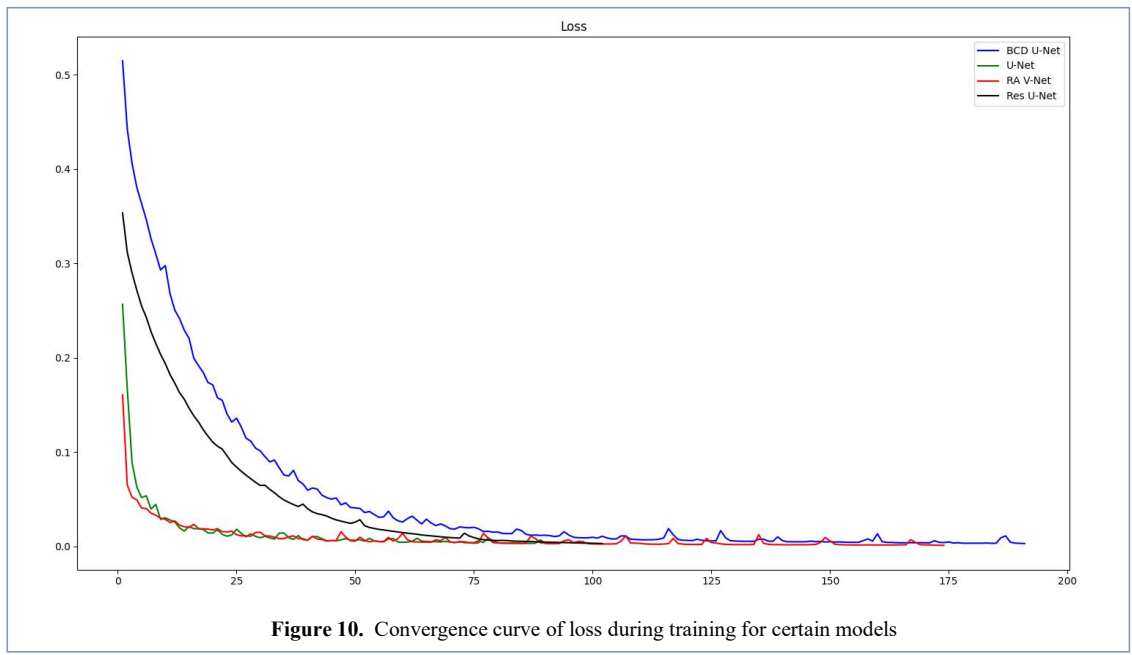

**Figure 10.** Convergence curve of loss during training for certain models

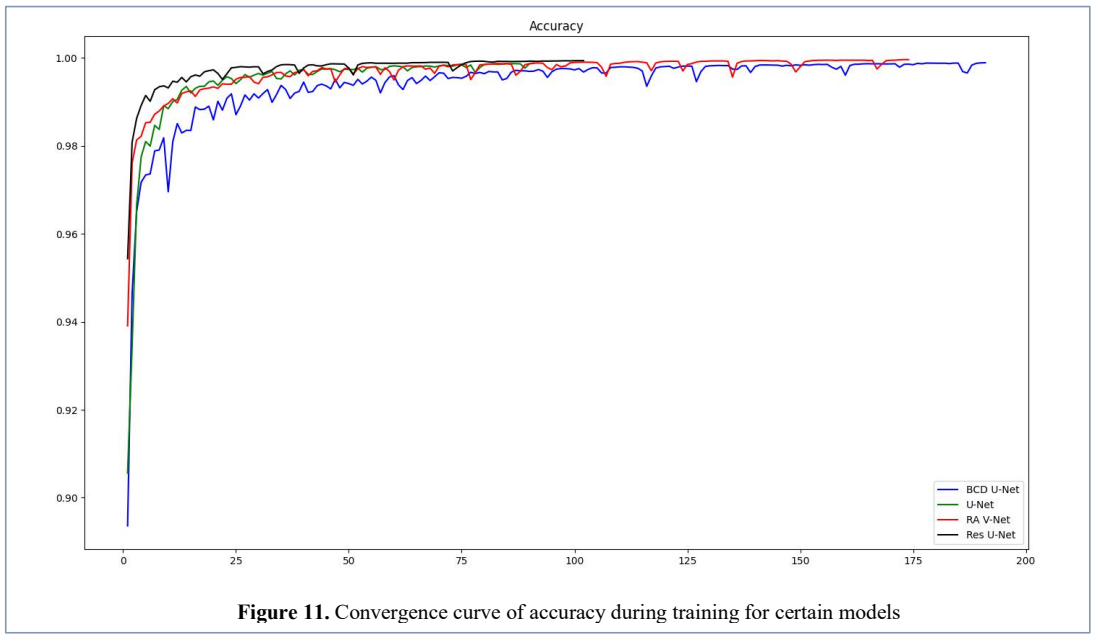

**Figure 11.** Convergence curve of accuracy during training for certain models

From **Table 4, Figure 10** and **Figure 11**, after combining the innovations of the two modules and introducing channel attention, the performance metrics of the model have made a qualitative leap. Our proposed model has a very fast learning speed, and it can achieve the expected results in a very short time.

After analysis, we believe that the excellent convergence performance of the model is influenced by the CofRes Module and attention mechanism. The clever use of residual blocks and the arrangement of BN make the model have strong resistance to overfitting, and the features of the images can be better extracted and preserved.

The faster training speed is also attributed to the attention mechanism setting. After several rounds of training, the model has a higher weight for the region to be segmented, which in turn can find and segment the target region faster. The implementation of spatial attention mechanism provided by the LSTM memory gates. The channel attention, located at the bottom of the model, allows the model to find the channel containing more image information from many channels and apply weights. Both attention mechanisms assist in model segmentation, which slightly increases the computational effort but provides a significant improvement for the speed of the model in the long run.

Finally, here are the actual segmentation results on the datasets for ours and several other model.

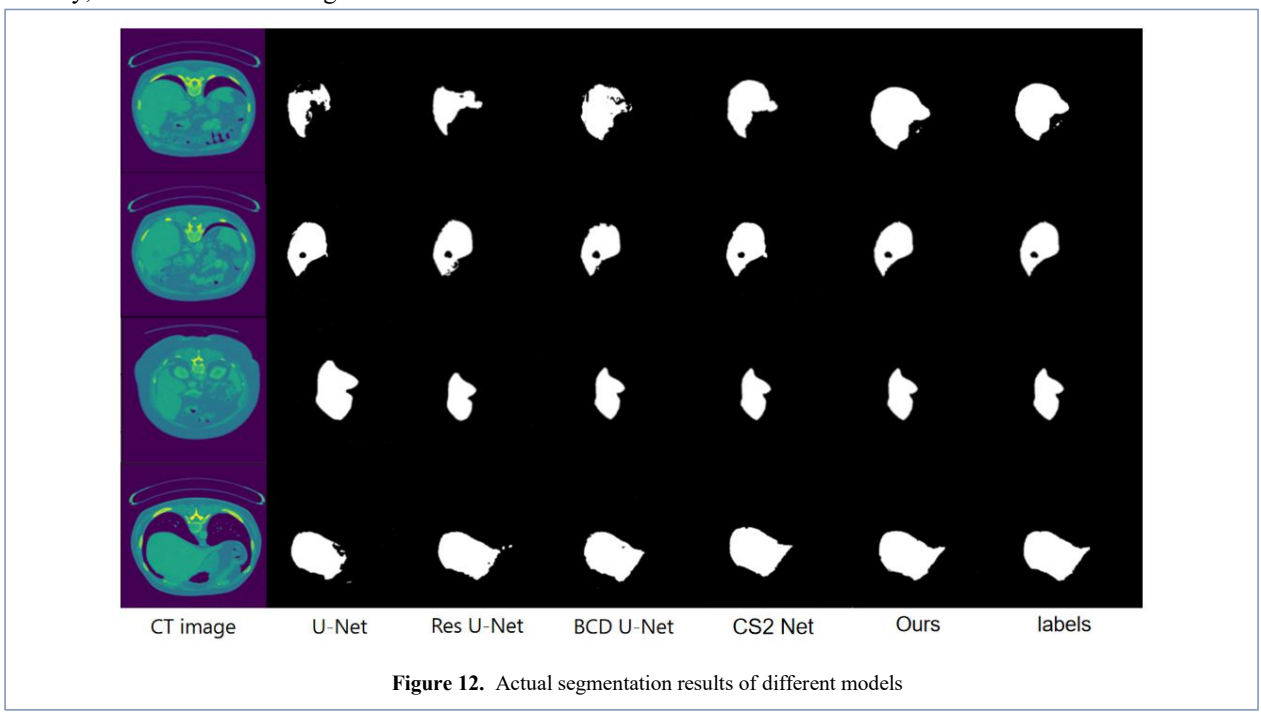

**Figure 12.** Actual segmentation results of different models

From the final segmentation results in Fig 12, each model does not segment the LiTS dataset poorly. However, some of the models are not fine and obvious enough for contour perception, like U-Net (Ronneberger *et al.*, 2015), BCD U-Net (Azad *et al.*, 2019). Some of the models are not good enough for details, like BCD U-Net (Azad *et al.*, 2019), CS2 Net (Mou *et al.*, 2021), there are tiny pixels with under-segmentation. In contrast, our proposed model performs very brightly in several test results. The two attention mechanisms focus on the overall contour and are combined with a deep CofRes Module to focus on extracting detailed features of the organ. AR Module reduces the computational effort and recovers the size of the underlying incoming feature matrix. The feature matrix from the encoding module and output from the upsapling block are fused and convoluted by the LSTM to increase the spatial correlation of pixs. It maximizes the retention of image features and boosts the weights for the regions to be segmented in the feature matrix, and has the functions of preventing gradient disappearance or gradient explosion, as well as preventing over-fitting. We use tedious ablation experiments to demonstrate the powerful application of our proposed two modules and the introduced channel attention mechanism for image feature extraction and size recovery. In the end, a strong segmentation performance was obtained during the actual segmentation.

## 4. Discussion

*4.1 Network architecture*

Since U-Net (Ronneberger *et al.*, 2015) was proposed in 2015, U-shaped networks have perennially dominated the field of deep learning automatic segmentation. The U-shaped structure consisting of an encoding module and a decoding module combined with a jump connection to preserve the original features represents a very efficient in image feature extraction. In the encoding module, the model extracts image features through the backbone network. The shallow features contain rich boundary information. And convolution extracts them well. U-Net continues convolution by pooling operations to reduce the image size after extracting the shallow boundary features. The purpose of this is to indirectly enhance the convolution kernel perceptual field by physical means. After the image size shrinkage is input to the backbone network, U-Net starts to extract the deep location features. At the bottom of U-Net is a densely connected layer that concatenates the encoding layers and decoding layers. The dense connection reduces the semantic difference between encoding and decoding. The decoding layer is mainly responsible for size recovery and aims to achieve end-to-end output, which is critical. The most important task in the decoding layer is to preserve the original features of the image. We tried to remove the jump connection of U-Net in our experiments, and the consequence of this is that the image features in the size recovery process are extremely different from the original features in the encoding layer. Accurate automatic segmentation cannot be accomplished.

Although U-Net has a bright performance and milestones, but still can not reach the application level. Therefore, a large number of researchers devoted to the optimization of U-Net, which has good performance but can only handle 2D medical images. V-Net (Milletari *et al.*, 2016) was proposed. V-Net can be used for segmentation of 3D medical images, which is more in line with the market demand. Unlike U-Net, V-Net removes the dense connectivity layer at the bottom of the network. It chooses to deliver the output of the backbone network directly to the decoding layer. One of the reasons is that the backbone network outperforms the original convolution layer in U-Net in terms of feature extraction development.

In recent years, other structures of networks are also trying to enter the field of medical image segmentation, like: Transformer (Matsoukas *et al.*, 2021), GAN. Transformer is a network structure consisting of a self-attentive mechanism. Transformer challenges network structures such as CNN and RNN with the advantages of powerful scaling, learning long distance dependence, etc. It has an unshakeable position in the field of NLP. And it did not stop here, in 2020, Transformer model was applied to image classification task for the first time and got close results to CNN model. Since then, many researches have started to try to apply the powerful ability of Transformer models to the field of computer vision, such as the Medical Transformer (Valanarasu *et al.*, 2021) with LoGo strategy. However, Transformer also has its drawbacks, such as weak local feature extraction ability and large dataset required for full training. GAN has been in full swing in the field of image generation. Although GAN is commonly used in the fields of denoising, reconstruction, and generation, CGAN also has potential in the field of medical image segmentation.

*4.2 Residuals module*

Since its introduction by He, the residual module (He *et al.*, 2016) has been making its presence felt in various fields with its powerful performance. It has been experimentally demonstrated that networks with intra-block connections have stronger ability to fit high-dimensional data. It provides powerful feature extraction capability for the model. For example, the performance of Res U-Net is significantly higher than that of U Net. The two aforementioned networks are identical except for the encoding module. This proves the usefulness of Res Net. In addition, using Res Net as the encoding block allows the network layers to be even very deep without the problem of gradient disappearance or gradient explosion.

Recently, researchers have started to focus on innovations for residual blocks. UNet3+, a Res2Net is proposed, based on the innovation proposed by Res Net. The number of parameters is reduced by a novel structure, and the accuracy of feature extraction is improved. In our experiments, the performance of our proposed CofRes Module is compared with Res Net and Res2Net using our proposed CofRes Module, and the performance of our proposed and Res2Net is approximate, each has its own bias, and both have significantly higher performance than Res Net. Res2Net is biased towards accracy while our CofRes Module focuses on sensitivity, both are almost the same in terms of F1 Score and Jaccard Similarity Score.

It is clear that the performance optimization of the encoding module is also crucial for the network as a whole.

*4.3 Attention mechanisms*

The role of the attention mechanism is to make the computer learn the pattern of the general distribution of data during the learning process. It allows the computer to have a region to focus, when receiving new image information. If this region is exactly the region to be segmented, then the training of this attention mechanism is successful. This attention mechanism automatically increases the weight of some regions after receiving the image to achieve the focused attention. Meanwhile, the parts whose weights are not increased are ignored in the segmentation process. The above description is the spatial attention mechanism, which is to learn the data distribution from the spatial perspective of the image. Besides, there are channel attention mechanism, soft attention mechanism and self-attention mechanism.

Channel attention is often added at the bottom of the model, where a large number of channels are stacked. It is done by inner-producting the feature matrix itself to obtain the weight matrix of semantic information embedded within each channel, and then multiplying it with the original feature matrix to do the purpose of weighting a specific channel. The other channel attention is similar.

Transformer consisted by self-attentive mechanism (Vaswani *et al.*, 2017). It is a variation of the attention mechanism, which is less dependent on external information and better at capturing the internal relevance of data or features. It is most commonly used with NLP, in textual applications, mainly to solve long-distance dependency problems by computing the interactions between words. Self-attention mechanism is also slowly influencing the field of computer vision in recent years.

From the above description, it can be seen that the attention mechanism, especially the spatial attention mechanism, enhances the model a lot. Therefore, in the next research, it is the current demand to design the attention mechanism with small operation and small number of parameters.

## 5. Conclusion

This simulation experiment demonstrates the difference in performance of deep learning models when applied to medical image segmentation and finds several directions to optimize the automated segmentation of liver, such as: improvement of residual blocks, combined application of multiple attention mechanisms. After improving the coding module, the performance of our model improves 0.085 over U-Net on DSC and 0.07 on JSC. 0.06 over Res U-Net on DSC. Experiments show that our proposed CofRes Module has stronger data fitting ability and feature extraction ability than Res Net. After improving the decoding module, our model improves 0.03 in DSC and 0.01 in JSC over BCD U-Net. This proves that our proposed AR Module combined with CA Module is effective for feature extraction and feature retention. After combining all the innovations, the proposed RA V-Net is in the leading position in deep learning liver segmentation models. The high performance segmentation provides for further identification of the lesion area and surgery by the surgeon.

Journal **XX** (XXXX) XXXXXX　　　　　　　　　　　　　　　　　　　　　　　　　　　　　　　　Author *et al*